\documentclass[letterpaper,twocolumn,american,showpacs,pra,aps,superscriptaddress,eqsecnum]{revtex4}
\usepackage[T1]{fontenc}
\usepackage[latin1]{inputenc}
\usepackage{bm}
\usepackage{amsmath}
\usepackage{babel}
\usepackage{graphicx}
\usepackage{amssymb}
\makeatletter
\usepackage{pifont}
\makeatother

\begin{document}

\title{Entanglement in $SO(3)$-invariant bipartite quantum systems}

\author{Heinz-Peter Breuer}

\email{breuer@physik.uni-freiburg.de}

\affiliation{Physikalisches Institut, Universit\"at Freiburg,
             Hermann-Herder-Str.~3, D-79104 Freiburg, Germany}

\date{\today}

\begin{abstract}
The structure of the state spaces of bipartite $N\otimes N$
quantum systems which are invariant under product representations
of the group $SO(3)$ of three-dimensional proper rotations is
analyzed. The subsystems represent particles of arbitrary spin $j$
which transform according to an irreducible representation of the
rotation group. A positive map $\vartheta$ is introduced which
describes the time reversal symmetry of the local states and which
is unitarily equivalent to the transposition of matrices. It is
shown that the partial time reversal transformation
$\vartheta_2=I\otimes \vartheta$ acting on the composite system
can be expressed in terms of the invariant $6$-$j$ symbols
introduced by Wigner into the quantum theory of angular momentum.
This fact enables a complete geometrical construction of the
manifold of states with positive partial transposition and of the
sets of separable and entangled states of $4\otimes 4$ systems.
The separable states are shown to form a three-dimensional prism
and a three-dimensional manifold of bound entangled states is
identified. A positive maps is obtained which yields, together
with the time reversal, a necessary and sufficient condition for
the separability of states of $4\otimes 4$ systems. The relations
to the reduction criterion and to the recently proposed cross norm
criterion for separability are discussed.
\end{abstract}


\maketitle

\section{Introduction}
It is one of the basic postulates of quantum mechanics that the
Hilbert space of states of a composite system is given by the
tensor product of the Hilbert spaces pertaining to its subsystems.
If a system is composed of two $N$-state systems with Hilbert
space ${\mathcal{H}}={\mathbb C}^N$, the mixed states of the total
system are represented by density matrices which act on the tensor
product ${\mathbb C}^N\otimes{\mathbb C}^N$. A state of such an
$N\otimes N$ system is said to be {\textit{separable}} or
{\textit{classically correlated}} if it can be generated by mixing
with certain probabilities an ensemble of states which describe
statistically independent subsystems \cite{WERNER}. States which
cannot be represented in this way are called
{\textit{inseparable}} or {\textit{entangled}}. The
characterization, classification and quantification of mixed-state
entanglement and the development of explicit necessary and
sufficient separability criteria turn out to be an extremely hard
problem \cite{ECKERT}. The solution of this problem could have
far-reaching consequences for fundamental questions of quantum
mechanics and computational complexity theory \cite{NIELSEN00} and
for applications in the theory of quantum information
\cite{ALBER,NIELSEN}.

A great simplification of the entanglement problem is obtained
though the introduction of symmetries
\cite{WERNER,VOLLBRECHT,BRACKEN}. By the requirement of invariance
under certain groups of symmetry transformations one restricts the
set of all states to a low-dimensional manifold of invariant
states and one may hope to get a tractable problem which is
solvable with the help of group theoretical and algebraic methods.
A prominent example is given by the one-parameter family of the
Werner states \cite{WERNER} which results from the requirement of
invariance under all product transformations of the form $U\otimes
U$, where $U$ varies over the full group of unitary $N\times N$
matrices. A further related example is the one-parameter family of
isotropic states \cite{HORODECKI99,RAINS} which are invariant
under all product unitaries $U\otimes U^*$, where $U^*$ denotes
the complex conjugation of $U$. Imposing the invariance under all
transformations of the form $O\otimes O$, where $O$ belongs to the
group of orthogonal $N\times N$ matrices, one obtains the
two-dimensional manifold of orthogonally invariant states
\cite{VOLLBRECHT}. It is clear that the larger the symmetry group
the smaller is the remaining space of invariant states and the
easier should be the analysis of its structure. In fact, the
problem of the explicit determination of the separable states
under symmetry requirements can be solved completely for the
examples given above.

A physically natural symmetry group is the group $SO(3)$ of proper
rotations in three dimensions. The underlying assumption is that
the states of the subsystems transform according to an
$N=(2j+1)$-dimensional irreducible representation of the rotation
group which corresponds to a fixed angular momentum $j$. The
subsystems thus behave under rotations as particles with a certain
spin $j$. The rotation group then operates through a reducible
product representation on the states of the composite system. Any
$SO(3)$-invariant state can be decomposed into a sum of
projections onto the irreducible subspaces belonging to the
various eigenvalues $J=0,1,\ldots,2j$ of the total angular
momentum operator. This shows that the rotationally invariant
states form an $(N-1)$-dimensional manifold. The requirement of
$SO(3)$ invariance reduces the full $(N^4-1)$-dimensional space of
all mixed states of an $N\otimes N$ system to an
$(N-1)$-dimensional space of invariant states.

The invariance under $SO(3)$ transformations represents in general
a much smaller symmetry than those of the examples given above.
For example, the manifolds of the Werner states and of the
isotropic states can be embedded into the set of rotationally
invariant states. These examples are thus special cases of the
$SO(3)$ symmetry.

The problem of mixed-state entanglement in $SO(3)$-invariant
bipartite systems will be analyzed in this paper. We find that the
state spaces exhibit an interesting convex structure and several
important phenomena as the emergence of non-decomposable positive
maps \cite{WORONOWICZ} and bound entanglement \cite{HORODECKI98}.
The physical significance of the $SO(3)$ symmetry derives from the
fact that any rotationally invariant state can be produced from
the maximally entangled angular momentum singlet state $J=0$
through the application of an isotropic dynamical map which
operates locally on the subsystems. The set of $SO(3)$-invariant
states is thus identical to the set of states which results from
interactions of the singlet state with noisy isotropic
environments.

A powerful tool in studies of entanglement is the operation of
taking the partial transposition of states. The requirement of
positive partial transposition (PPT) represents a strong necessary
condition for the separability of states, known as the Peres
criterion \cite{PERES}. A conceptually simple but crucial point of
the present investigation consists in the replacement of the
transposition by another unitarily equivalent operation which is
identical to the time reversal transformation of particles with
spin $j$. It is known that the anti-unitary operation of the time
reversal commutes with the representations of the rotation group.
This fact allows one to characterize the partial transposition by
means of invariant quantities which are directly related to
Wigner's $6$-$j$ symbols \cite{WIGNER}.

The $6$-$j$ symbols arise in the transformation between different
coupling schemes for the addition of angular momenta
\cite{EDMONDS}. They can be expressed as invariant sums over
products of vector-coupling (Clebsch-Gordan) coefficients. Thus we
find a close connection between the partial transposition, the
time reversal symmetry and certain group-theoretical invariants
built out of the vector-coupling coefficients. It will be
demonstrated here that this connection to group-theoretical
concepts leads to important implications on the entanglement
structure of the state space.

The content of the paper can be summarized as follows. In
Sec.~\ref{DD-INVAR} we briefly recall some facts from the
representation theory of the rotation group and introduce an
appropriate parametrization of the set of rotationally invariant
states. The partial transposition and the corresponding
transformation of the partial time reversal are discussed in
Sec.~\ref{MAPS}. It is shown that this transformation preserves
the rotational invariance of states and its relation to the Wigner
$6$-$j$ symbols is derived.

These results are used in Secs.~\ref{PPT-STATES} and \ref{SEP} to
develop a geometric representation of the sets of the PPT states
and of the separable states in the cases $N=2$, $3$ and $4$. Most
importantly, in the case $N=4$ we find that the set of separable
states is isomorphic to a three-dimensional prism, i.~e., to a
polyhedron which is bounded by three squares and two triangles. We
further identify a three-dimensional manifold of bound entangled
states with positive partial transposition.

Finally, Sec.~\ref{CONCLU} contains a discussion of the results
and a number of conclusions which can be drawn from the present
investigation. In particular, we construct a positive map which
yields, together with the time reversal, a necessary and
sufficient condition for separability in the case of $4\otimes 4$
systems. Moreover, we discuss the relations to two further
criteria of separability, namely the reduction criterion and the
cross norm criterion.

\section{The set of $SO(3)$-invariant states}\label{DD-INVAR}

\subsection{Representations of the rotation group}\label{SO3}
We consider a bipartite quantum system whose local parts are
$N$-state systems with corresponding state space
${\mathcal{H}}={\mathbb C}^N$. The Hilbert space of the composite
system is given by the tensor product space
${\mathcal{H}}\otimes{\mathcal{H}}$. The local state spaces are
regarded as angular momentum manifolds corresponding to a certain
eigenvalue of the square of the angular momentum operator
$\hat{\bm{j}}$. Thus, the state space ${\mathcal{H}}$ is spanned
by a fixed orthonormal basis of $N=(2j+1)$ angular momentum
eigenvectors $|jm\rangle$, where $m=-j,-j+1,\ldots,+j$. As usual
we have the eigenvector relations
$\hat{\bm{j}}^2|jm\rangle=j(j+1)|jm\rangle$ and
$\hat{j}_3|jm\rangle=m|jm\rangle$. Note that $j$ can take on
integer or half-integer values,
$j=\frac{1}{2},1,\frac{3}{2},\ldots$, such that $N=2,3,4,\ldots$

The group of proper rotations in three dimension is denoted by
$SO(3)$. This is the group of orthogonal $3\times 3$ matrices with
determinant $1$. An irreducible representation of this group on
the state space ${\mathcal{H}}$ is obtained in the standard way:
Given a rotation $R \in SO(3)$ the corresponding transformation of
state vectors is provided by the unitary matrix
\begin{equation} \label{REP-ROT}
 D(R) = D(n_1,n_2,n_3) = \exp \left(-i\bm{n}\cdot\hat{\bm{j}}\right).
\end{equation}
The rotation $R$ is characterized here by the vector
${\bm{n}}=(n_1,n_2,n_3)$, i.~e., $R$ is the rotation about the
axis given by $\bm{n}$ by the angle $|\bm{n}|$ (in a right-handed
sense). It should be mentioned that Eq.~(\ref{REP-ROT}) generally
yields a two-valued representation of the rotation group: For
half-integer $j$ one obtains two unitary matrices which represent
a given rotation $R$ and which differ in sign.

\subsection{Rotational invariance of bipartite systems}
The representation (\ref{REP-ROT}) leads to a representation of
the rotation group on the tensor product space
${\mathcal{H}}\otimes{\mathcal{H}}$ of the bipartite system. If
$\rho$ is an operator acting on the tensor product space, a
rotation $R$ carried out on both parts of the composite system
leads to the transformed operator $\rho'=[D(R) \otimes
D(R)]\rho[D(R)\otimes D(R)]^{\dagger}$. An operator $\rho$ is said
to be rotationally invariant or $SO(3)$-invariant if it is
invariant under all such transformation, that is, if the relation
\begin{equation} \label{TENSOR-REP}
 [D(R) \otimes D(R)] \rho [D(R) \otimes D(R)]^{\dagger} = \rho
\end{equation}
holds for all $R\in SO(3)$.

A state of the bipartite system is given by a density matrix
$\rho$ satisfying $\rho \geq 0$ and ${\mathrm{tr}}\rho=1$. The set
of all states $\rho$ which fulfill the invariance requirement
(\ref{TENSOR-REP}) will be denoted by $S$. It is clear that $S$ is
a convex subset of the set of all states of the bipartite system.

The angular momentum operator of the composite system is given by
$\hat{\bm{J}}=\hat{\bm{j}}\otimes I + I \otimes \hat{\bm{j}}$,
where $I$ denotes the unit matrix. The components of
$\hat{\bm{J}}$ are the generators of the product representation
and the requirement of rotational invariance is equivalent to the
statement that $\rho$ commutes with all components of
$\hat{\bm{J}}$.

The product representation $D(R)\otimes D(R)$ is obviously
reducible. Its decomposition into a sum of irreducible
representations is a standard subject of quantum mechanics. One
introduces an orthonormal basis in
${\mathcal{H}}\otimes{\mathcal{H}}$ which consists of the common
eigenvectors $|JM\rangle$ of $\hat{\bm{J}}^2$ and $\hat{J}_3$
corresponding to the eigenvalues $J(J+1)$ and $M$, respectively,
where $J=0,1,\ldots,2j$ and $M=-J,-J+1,\ldots,+J$. The
$(2J+1)$-dimensional space which is spanned by the basis vectors
$|JM\rangle$ with a fixed $J$ is an invariant and irreducible
subspace of the tensor product representation.

The set $S$ of rotationally invariant operators can now easily be
characterized. To this end, we introduce projection operators
\begin{equation} \label{DEF-PJ}
 P_J = \sum_{M=-J}^{+J} |JM\rangle\langle JM|
\end{equation}
which project onto the subspaces belonging to a fixed $J$. From
the irreducibility of the representation within these subspaces
one concludes with the help of Schur's lemma that any rotationally
invariant operator $\rho$ can be written as a linear combination
of the projections:
\begin{equation} \label{D-REP}
 \rho = \frac{1}{N} \sum_{J=0}^{2j} \frac{\alpha_J}{\sqrt{2J+1}} P_J.
\end{equation}
Here, the $\alpha_J$ are c-numbers and we have introduced
normalization factors $(N\sqrt{2J+1})^{-1}$. It will be seen in
Sec.~\ref{PROP-THETA} that this choice of normalization factors
leads to highly symmetric transformation properties of the
parameter space. For $\rho$ to be Hermitian the $\alpha_J$ must of
course be real. Equation (\ref{D-REP}) then corresponds to the
spectral decomposition of $\rho$. If $\rho$ is a density matrix
the $\alpha_J$ are real and positive, $\alpha_J \geq 0$. On using
${\mathrm{tr}}P_J = 2J+1$, the normalization condition takes the
form
\begin{equation} \label{NORMIERUNG}
 {\mathrm{tr}}\rho = \sum_{J=0}^{2j} \frac{\sqrt{2J+1}}{N} \alpha_J
 = 1.
\end{equation}

For example, setting $\alpha_0=N$ and $\alpha_J=0$ for
$J=1,2,\ldots,2j$ we get $\rho=P_0=|00\rangle\langle 00|$, i.~e.,
the projection onto the angular momentum singlet state
\begin{equation}
 |00\rangle = \frac{1}{\sqrt{N}}\sum_{m=-j}^{+j} (-1)^{j-m}
 |j,m\rangle \otimes |j,-m\rangle.
\end{equation}
This state is the only pure state in $S$ and it is maximally
entangled (the quantity $\frac{\alpha_0}{N}$ is the singlet
fraction). Using the completeness of the projections $P_J$ one
concludes that the state corresponding to
$\alpha_J=\sqrt{2J+1}/N$, $J=0,1,\ldots,2j$, is the separable
state $\rho=\frac{1}{N^2}I\otimes I$ of maximal entropy.

It follows from the irreducibility of the representation $D(R)$
that for any $SO(3)$-invariant state $\rho$ the reduced density
matrices $\rho^{(1)}={\mathrm{tr}}_2\rho$ and
$\rho^{(2)}={\mathrm{tr}}_1\rho$ of the subsystems, given by the
partial traces ${\mathrm{tr}}_2$ and ${\mathrm{tr}}_1$, are
proportional to the identity $I$. The reduced density matrices
obtained from a rotationally invariant state thus describe states
of maximal disorder.

Summarizing, by means of Eq.~(\ref{D-REP}) any rotationally
invariant Hermitian operator is uniquely characterized by $N$ real
parameters $\alpha_J$. We can therefore identify the set of all
such operators with the set of points
\begin{equation}
 {\bm{\alpha}} = \left( \begin{array}{c}
 \alpha_0 \\ \alpha_1 \\ . \\ . \\ \alpha_{2j} \end{array}
 \right) \in {\mathbb R}^N
\end{equation}
in an $N$-dimensional parameter space ${\mathbb R}^N$. The set of
points $\bm{\alpha}$ in this space satisfying $\alpha_J \geq 0$
and the normalization condition (\ref{NORMIERUNG}) then describes
the set $S$ of rotationally invariant density matrices. In
geometrical terms $S$ represents an $(N-1)$-dimensional simplex.
For instance, $S$ is a line for $N=2$, a triangle for $N=3$, and a
tetrahedron for $N=4$. These examples will be discussed in
Secs.~\ref{EXAMPLES-1} and \ref{EXAMPLES-2}.

\section{Positive maps and rotational invariance}\label{MAPS}

\subsection{Partial transposition}
Given an operator $B$ on ${\mathcal{H}}$ the transposed operator
$TB = B^T$ is defined in terms of the local basis states
$|jm\rangle$ by means of $\langle jm|B^T|jm'\rangle \equiv \langle
jm'|B|jm\rangle$. Correspondingly, the partial transposition
$T_2=I\otimes T$ on the tensor product space is defined through
\begin{equation} \label{DEF-T2}
 T_2 (A\otimes B) = A \otimes TB = A \otimes B^T.
\end{equation}
The operation of taking the partial transpose plays an important
role in entanglement and quantum information theory. One reason
for this fact is that $T$ is a distinguished example of a map
which is positive but not completely positive
\cite{STINESPRING,CHOI1,CHOI2,WORONOWICZ,KRAUS1,KRAUS2}. This
means that $T$ takes positive operators on ${\mathcal{H}}$ to
positive operators on ${\mathcal{H}}$, while $T_2\rho$ need not be
positive for a positive operator $\rho$ on the tensor product
space ${\mathcal{H}}\otimes {\mathcal{H}}$.

Important information on the entanglement structure of states is
obtained by considering the action of positive but not completely
positive maps. An example is given by the Peres PPT criterion
according to which positivity under the partial transposition
$T_2$ is a necessary condition for separability \cite{PERES}. An
important general characterization has been developed by the
Horodecki's \cite{HORODECKI96a}: A necessary and sufficient
condition for a state $\rho$ to be separable is that the operator
$(I\otimes \Phi)\rho$ is positive for any positive map $\Phi$.
This condition does however not lead to a simple operational
criterion for separability since we have no general structural
characterization of positive maps, as it exists for completely
positive maps in the form of the Kraus-Stinespring representation
\cite{STINESPRING,KRAUS1,KRAUS2}.

\subsection{$\vartheta_2$-transformation}\label{CONSTR-THETA}
If $\rho$ is a rotationally invariant operator the partially
transposed operator $T_2\rho$ is generally not invariant under
rotations. It can be shown that, instead, $T_2\rho$ is invariant
under transformations of the form $D(R)\otimes D(R)^*$, where
$D(R)^*$ is the matrix obtained from $D(R)$ by complex conjugation
of its elements, that is $D(R)^*=D(R)^{\dagger T}$. Throughout
this paper $T$ denotes the transposition, $\dagger$ the adjoint
and $\ast$ the element-wise complex conjugation of a matrix.

In the present investigation we shall utilize a map which is
unitarily equivalent to the partial transposition, but which does
map rotationally invariant operators to rotationally invariant
operators. This map will be denoted by $\vartheta_2$. By analogy
to Eq.~(\ref{DEF-T2}), $\vartheta_2$ is taken to be of the form
\begin{equation} \label{DEF-VARTHETA}
 \vartheta_2(A\otimes B) = A\otimes \vartheta B
 = A\otimes V B^T V^{\dagger}
\end{equation}
with some fixed unitary matrix $V$. Hence, $\vartheta_2=I\otimes
\vartheta$ is the partial transposition $T_2$ followed by a local
unitary transformation acting on the second part of the bipartite
system, that is, we have $\vartheta_2\rho=(I\otimes
V)T_2\rho(I\otimes V)^{\dagger}$. Since the maps $\vartheta_2$ and
$T_2$ are unitarily equivalent a state $\rho$ is obviously
positive under $\vartheta_2$ if and only if it is positive under
$T_2$.

The unitary matrix $V$ will be determined from the condition that
$\vartheta_2$ preserves the rotational invariance of operators,
i.~e., if $\rho$ is any invariant operator satisfying
Eq.~(\ref{TENSOR-REP}) we demand that the transformed operator
$\vartheta_2\rho$ is again invariant:
\begin{equation} \label{INVAR-THETA}
 [D(R) \otimes D(R)] \vartheta_2\rho [D(R) \otimes D(R)]^{\dagger}
 = \vartheta_2\rho.
\end{equation}
This requirement is obviously satisfied if the map $\vartheta$
commutes with all rotations, that is, if the relation
\begin{equation} \label{INVAR-THETA-1}
 \vartheta[D(R) B D(R)^{\dagger}] = D(R) (\vartheta B) D(R)^{\dagger}
\end{equation}
holds true for all operators $B$ on ${\mathcal{H}}$ and all $R\in
SO(3)$. By use of the definition of $\vartheta$ given by
Eq.~(\ref{DEF-VARTHETA}) one can write Eq.~(\ref{INVAR-THETA-1})
as
\begin{equation} \label{INVAR-THETA-2}
 VD(R)^*B^T[VD(R)^*]^{\dagger} = D(R)VB^T[D(R)V]^{\dagger}.
\end{equation}
This equation is fulfilled if $VD(R)^*=D(R)V$. Thus we see that
the rotational invariance of $\vartheta_2\rho$ follows from the
rotational invariance of $\rho$ provided we can find a fixed
unitary matrix $V$ such that
\begin{equation} \label{INVAR-THETA-3}
 VD(R)^*V^{\dagger} = D(R)
\end{equation}
for all $R\in SO(3)$.

To obtain a unitary matrix $V$ satisfying
Eq.~(\ref{INVAR-THETA-3}) we employ specific properties of the
representations of the rotation group. As in Eq.~(\ref{REP-ROT}),
let $D(R)$ be the representation of the rotation $R$ about an axis
$\bm{n}=(n_1,n_2,n_3)$ by an angle $|\bm{n}|$. The complex
conjugation of the elements of $D(R)$ then yields the matrix
\begin{eqnarray} \label{REP-ROT*}
 D(R)^* &=& \exp \left(+i\bm{n}\cdot\hat{\bm{j}}^T\right)
 \nonumber \\
 &=& \exp \left(-i[-n_1\hat{j}_1 + n_2\hat{j}_2 - n_3\hat{j}_3
 ]\right) \nonumber \\
 &=& \exp \left(-i\bm{n}'\cdot\hat{\bm{j}}\right).
\end{eqnarray}
Here we use the fact that in the local basis $|jm\rangle$ the
transposed components of the angular momentum operator are given
by $\hat{j}_1^T=\hat{j}_1$, $\hat{j}_2^T=-\hat{j}_2$ and
$\hat{j}_3^T=\hat{j}_3$. Thus, $D(R)^*$ represents the rotation
about the axis ${\bm{n}}'=(-n_1,n_2,-n_3)$ which is obtained from
${\bm{n}}$ through a rotation by $\pi$ about the $x_2$-axis. To
transform from $D(R)^{\ast}$ to $D(R)$ we therefore define $V$ to
be the matrix representing a $\pi$-rotation about the $x_2$-axis.
Using the notation introduced in Eq.~(\ref{REP-ROT}) we write
\begin{equation} \label{DEF-V}
 V \equiv D(0,\pi,0).
\end{equation}
Explicitly the matrix elements of $V$ are given by
\begin{equation} \label{ELEMENTS-V}
 \langle jm'|V|jm\rangle=(-1)^{j-m}\delta_{m',-m}.
\end{equation}
Hence, $V$ is real and we have $V^T=V^{\dagger}=V^{-1}$.

Equation (\ref{DEF-V}) yields
\begin{equation}
 V \left( \bm{n}' \cdot \hat{\bm{j}} \right) V^{\dagger}
 = \bm{n} \cdot \hat{\bm{j}},
\end{equation}
which, by use of Eq.~(\ref{REP-ROT*}), immediately leads to the
desired relation (\ref{INVAR-THETA-3}). We conclude that the map
$\vartheta_2$ defined by Eqs.~(\ref{DEF-VARTHETA}) and
(\ref{DEF-V}) preserves the rotational invariance of operators.
The advantage of this formulation is that $\vartheta_2$, by
contrast to $T_2$, maps the set of rotationally invariant
Hermitian operators onto itself and can be expressed as a simple
linear transformation of the parameters $\alpha_J$. This
transformation will be determined in Sec.~\ref{6j}.

\subsection{Time reversal symmetry}\label{TIME-REVERSAL}
The transposition $T$ is closely connected to the operation of
reversing the direction of motion, i.~e., to the symmetry
transformation of time reversal
\cite{WORONOWICZ,SANPERA,HORODECKI98}. We demonstrate that, in
fact, it is the map $\vartheta$ introduced in
Eq.~(\ref{DEF-VARTHETA}) which describes the time reversal of
particles with spin $j$.

We have seen in the preceding subsection that $T$ changes the sign
of $\hat{j}_2$ and leaves $\hat{j}_1$ and $\hat{j}_3$ unchanged,
while the unitary operator $V$ (representing a $\pi$-rotation
about the $x_2$-axis) changes the signs of $\hat{j}_1$ and
$\hat{j}_3$ and leaves $\hat{j}_2$ unchanged. Hence, we have
$\vartheta \hat{\bm{j}}=V\hat{\bm{j}}^TV^{\dagger}=-\hat{\bm{j}}$.
This shows that the map $\vartheta$ describes the behaviour of the
angular momentum operator under time reversal.

It is known from Wigner's representation theorem \cite{WIGNER}
that the time reversal symmetry must be represented in terms of an
anti-unitary operator. Indeed, we can express the action of
$\vartheta$ by means of an anti-unitary operator $\tau$ through
\begin{equation} \label{THETA-TAU}
 \vartheta B = \tau B^{\dagger} \tau^{-1}.
\end{equation}
The operator $\tau=V\tau_0$ is composed of the unitary
transformation $V$ introduced above and of the anti-unitary
transformation $\tau_0$ which is given by the complex conjugation
of the amplitudes in the basis $|jm\rangle$:
\begin{equation} \label{TAU-0}
 |\varphi\rangle = \sum_m c_m |jm\rangle \mapsto
 \tau_0|\varphi\rangle = \sum_m c^{\ast}_m |jm\rangle.
\end{equation}
Thus, by virtue of Eq.~(\ref{ELEMENTS-V}) we have
\begin{equation} \label{TAU}
 |\varphi\rangle = \sum_m c_m |jm\rangle \mapsto
 \tau|\varphi\rangle = \sum_m c^{\ast}_m (-1)^{j-m}|j,-m\rangle.
\end{equation}
This transformation expresses the well-known behaviour of spin-$j$
particles under time reversal. For example, in the case $N=2$
($j=\frac{1}{2}$ and $m=\pm\frac{1}{2}$) Eq.~(\ref{ELEMENTS-V})
leads to $V=-i\sigma_2$, where $\sigma_2$ is a Pauli matrix. The
transformation $\tau$ thus consist of the complex conjugation and
of the unitary transformation given by the matrix $-i\sigma_2$,
which precisely corresponds to the time reversal transformation of
a spin-$\frac{1}{2}$ particle.

In view of these results the map $\vartheta_2=I\otimes\vartheta$
may be interpreted as a {\textit{partial time reversal}} of the
composite system. The fact that $\vartheta$ is not completely
positive means that the operation of time reversal, when carried
out only on a subsystem, does in general not lead to a physically
legitimate state \cite{LAHTI}.

\subsection{Properties of the map $\vartheta_2$}\label{PROP-THETA}
The properties of the transformations $\vartheta$ and
$\vartheta_2$ are of course very similar to those of the
transposition $T$ and of the partial transposition $T_2$,
respectively. In particular, $\vartheta$ is a positive (but not
completely positive) map, i.~e., $B \geq 0$ implies that
$\vartheta B \geq 0$. Moreover, $\vartheta$ preserves the trace,
${\mathrm{tr}}\{\vartheta B\}={\mathrm{tr}} B$, and the unit
matrix, $\vartheta I = I$.

It follows from Eq.~(\ref{ELEMENTS-V}) that
\begin{equation} \label{PROP-V}
 V^2 = (-1)^{2j} I.
\end{equation}
This equation illustrates the two-valuedness of the
representation: $V^2=D(0,\pi,0)^2$ represents a rotation by $2\pi$
(i.~e., the identity in $SO(3)$) and is equal to $-I$ for
half-integer $j$. Equation (\ref{PROP-V}) leads to the conclusion
that, like $T_2$, the map $\vartheta_2$ is an involution which
means that $\vartheta_2^2 = I \otimes \vartheta^2$ is equal to the
identity map. In fact, employing Eq.~(\ref{PROP-V}) we obtain for
any operator $B$ on ${\mathcal{H}}$:
\begin{equation}
 \vartheta(\vartheta B) = V(VB^TV^{\dagger})^TV^{\dagger}
 = VVBV^{\dagger}V^{\dagger} = B.
\end{equation}
Another property which will be important below is that
$\vartheta_2$ is selfadjoint with respect to the Hilbert-Schmidt
inner product, i.~e., we have
\begin{equation} \label{SELFADJOINT}
 {\mathrm{tr}}\left\{ X^{\dagger} (\vartheta_2 Y) \right\}
 =  {\mathrm{tr}}\left\{ (\vartheta_2 X)^{\dagger} Y \right\}
\end{equation}
for all operators $X$ and $Y$ on the tensor product space. This
property derives from the corresponding property of the map
$\vartheta$. Namely, for any two operators $A$ and $B$ on
${\mathcal{H}}$ we have according to Eq.~(\ref{DEF-VARTHETA}):
\begin{eqnarray}
 {\mathrm{tr}}\{A^{\dagger}(\vartheta B)\}
 &=& {\mathrm{tr}}\{A^{\dagger}VB^TV^{\dagger}\} \nonumber \\
 &=& {\mathrm{tr}}\{(V^TA^TV^{\dagger T})^{\dagger}B\} \nonumber \\
 &=& (-1)^{4j} {\mathrm{tr}}\{(VA^TV^{\dagger})^{\dagger}B\}
 \nonumber \\
 &=& {\mathrm{tr}}\{(\vartheta A)^{\dagger}B\}.
\end{eqnarray}
Note that we have used here that $(-1)^{4j}=1$ for integer and
half-integer $j$, and that $V^T=V^{-1}=(-1)^{2j}V$, which follows
from Eq.~(\ref{PROP-V}).

Since $\vartheta$ is not completely positive the operator
$\rho'=\vartheta_2\rho$ need not be positive for a positive
$\rho$. It is, however, invariant under rotations and can be
represented in the form (\ref{D-REP}). To determine the action of
$\vartheta_2$ on the $\alpha_J$-parameters we therefore write:
\begin{eqnarray}
 \rho' &=& \frac{1}{N} \sum_{K=0}^{2j} \frac{\alpha'_K}{\sqrt{2K+1}} P_K
 \nonumber \\
 &=& \vartheta_2\rho = \frac{1}{N} \sum_{K=0}^{2j}
 \frac{\alpha_K}{\sqrt{2K+1}} \vartheta_2 P_K,
\end{eqnarray}
where the parameters $\alpha_K$ correspond to $\rho$ and the
$\alpha_K'$ correspond to $\rho'$. We multiply this equation by
$P_J$ and take the trace using $P_JP_K=\delta_{JK}P_K$. This
yields a linear transformation from the parameters $\alpha_K$ to
the parameters $\alpha'_K$. Using matrix notation we find
\begin{equation} \label{TRAFO-THETA}
 \bm{\alpha}' = \Theta \bm{\alpha},
\end{equation}
where we have introduced a matrix $\Theta$ with elements
\begin{equation} \label{DEF-THETA}
 \Theta_{JK} = \frac{1}{\sqrt{(2J+1)(2K+1)}}
 {\mathrm{tr}}\{ P_J \vartheta_2 P_K\}.
\end{equation}
The map $\vartheta_2$ thus induces a linear transformation of the
parameter space which is given by the $N\times N$ matrix $\Theta$.

The matrix $\Theta$ is real symmetric, $\Theta^T=\Theta$, and
orthogonal, $\Theta^T\Theta=I$. The symmetry follows immediately
from definition (\ref{DEF-THETA}) and the property
(\ref{SELFADJOINT}). Since $\vartheta_2$ is an involution the
matrix $\Theta$ must also be an involution, that is $\Theta^2=I$.
Together with the symmetry of $\Theta$ we therefore have
$\Theta^T\Theta=\Theta^2=I$, which proves that $\Theta$ is
orthogonal.

\subsection{Relation to Wigner's $6$-$j$ symbols}\label{6j}
We derive a general expression for the elements of the matrix
$\Theta$. It will be shown that these elements are closely linked
to Wigner's $6$-$j$ symbols. To this end, we use
Eq.~(\ref{DEF-THETA}) as well as the definition (\ref{DEF-PJ}) of
the projections $P_J$ in terms of the eigenbasis $|JM\rangle$,
which gives
\begin{widetext}
\begin{equation} \label{DERIV-1}
 \Theta_{JK} = \frac{1}{\sqrt{(2J+1)(2K+1)}}
 \sum_{M=-J}^{+J} \sum_{Q=-K}^{+K}
 \langle JM|\vartheta_2\big( |KQ\rangle\langle KQ| \big) |JM\rangle.
\end{equation}
To evaluate the $\vartheta_2$-transformation in this expression we
insert complete sets of product basis states $|jmjm'\rangle$ to
get
\[
 \Theta_{JK} = \frac{1}{\sqrt{(2J+1)(2K+1)}}
 \sum_{M,Q} \sum_{m_1,m_2} \sum_{m_4,m_5}
 \langle JM|\vartheta_2\big( |m_1m_2\rangle\langle m_1m_2|KQ\rangle
 \langle KQ|m_4m_5\rangle\langle m_4m_5| \big) |JM\rangle.
\]
Here and in the following we shall frequently abbreviate
$|jm_1jm_2\rangle$ by $|m_1m_2\rangle$, etc. According to the
definition of $\vartheta_2$ [Eqs.~(\ref{DEF-VARTHETA}) and
(\ref{DEF-V})] and to Eq.~(\ref{ELEMENTS-V}) we have
\begin{eqnarray} \label{DERIV-3}
 \vartheta_2\big( |m_1m_2\rangle\langle m_4m_5| \big)
 &=& |m_1\rangle\langle m_4| \otimes
 V \left( |m_2\rangle\langle m_5| \right)^T V^{\dagger}
 = |m_1\rangle\langle m_4| \otimes V |m_5\rangle\langle m_2| V^{\dagger}
 \nonumber \\
 &=& |m_1\rangle\langle m_4| \otimes (-1)^{2j-m_2-m_5} |-m_5\rangle\langle -m_2|
 \nonumber \\
 &=& (-1)^{2j-m_2-m_5} |m_1,-m_5\rangle\langle m_4,-m_2|,
\end{eqnarray}
which leads to
\begin{equation} \label{DERIV-4}
 \Theta_{JK} = \frac{1}{\sqrt{(2J+1)(2K+1)}}
 \sum_{M,Q} \sum_{m_1,m_2} \sum_{m_4,m_5} (-1)^{2j-m_2-m_5}
 \langle JM|m_1,-m_5\rangle \langle m_1m_2|KQ\rangle
 \langle KQ|m_4m_5\rangle \langle m_4,-m_2|JM\rangle.
\end{equation}
\end{widetext}
The matrix elements in Eq.~(\ref{DERIV-4}) are vector-coupling
(Clebsch-Gordan) coefficients. Throughout the paper we adopt the
usual phase conventions for these quantities, as they are given,
e.~g., in Ref.~\cite{EDMONDS}.

To evaluate further Eq.~(\ref{DERIV-4}) it is convenient to employ
the $3$-$j$ symbols
\begin{equation} \label{DEF-3j}
 \left(\begin{array}{ccc}
 j_1 & j_2 & j_3 \\
 m_1 & m_2 & m_3
 \end{array}\right)
\end{equation}
introduced by Wigner. These quantities are known from the theory
of angular momentum coupling and are closely related to the
vector-coupling coefficients. Here, we have
\begin{equation} \label{DERIV-5}
 \langle m_1m_2|JM\rangle = (-1)^M \sqrt{2J+1}
 \left(\begin{array}{ccc}
 j & j & J \\
 m_1 & m_2 & -M
 \end{array}\right).
\end{equation}
The $3$-$j$ symbols have many symmetry properties. The symmetry to
be used here is given by
\begin{equation} \label{DERIV-6}
 \left(\begin{array}{ccc}
 j & j & J \\
 m_1 & m_2 & m_3
 \end{array}\right)
 = (-1)^{2j+J}
 \left(\begin{array}{ccc}
 j & j & J \\
 -m_1 & -m_2 & -m_3
 \end{array}\right).
\end{equation}
We further need the selection rules for the $3$-$j$ symbols,
namely that (\ref{DEF-3j}) is equal to zero for $m_1+m_2+m_3 \neq
0$.

We introduce the relations (\ref{DERIV-5}) into
Eq.~(\ref{DERIV-4}) which yields a sum over products of four
$3$-$j$ symbols. In the resulting expression we carry out the
following manipulations: (i) we interchange the summation indices
$m_2$ and $m_5$, (ii) we replace the summation index $m_1$ by
$-m_1$, (iii) we introduce the new notation $M\equiv m_3$,
$Q\equiv m_6$, and (iv) we employ the symmetry relation
(\ref{DERIV-6}) in the first and the third $3$-$j$ symbol. These
manipulations lead to
\begin{eqnarray} \label{DERIV-7}
 \Theta_{JK} &=& \sqrt{(2J+1)(2K+1)}\sum_{m_1,\ldots,m_6} \chi(m_i) \\
 &~& \times
 \left(\begin{array}{ccc}
 j & j & J \\
 m_1 & m_2 & m_3
 \end{array}\right)
 \left(\begin{array}{ccc}
 j & j & K \\
 -m_1 & m_5 & -m_6
 \end{array}\right)        \nonumber \\
 &~& \times
 \left(\begin{array}{ccc}
 j & j & K \\
 -m_4 & -m_2 & m_6
 \end{array}\right)
 \left(\begin{array}{ccc}
 j & j & J \\
 m_4 & -m_5 & -m_3
 \end{array}\right), \nonumber
\end{eqnarray}
where all sign factors have been collected in the quantity
\begin{equation} \label{DERIV-8}
 \chi(m_i) = (-1)^{2j-m_2-m_5+J+K}.
\end{equation}
Finally, we use the selection rules for the first and the third
$3$-$j$ symbol in Eq.~(\ref{DERIV-7}) which leads to
$m_1+m_2+m_3=0$ and $-m_4-m_2+m_6=0$. With the help of these
relations it is easy to show that the phase factor $\chi(m_i)$ may
be written as
\begin{eqnarray} \label{DERIV-9}
 \chi(m_i) &=&
 (-1)^{j+m_1}(-1)^{j+m_2}(-1)^{J+m_3} \nonumber \\
 &~& \times
 (-1)^{j+m_4}(-1)^{j+m_5}(-1)^{K+m_6}.
\end{eqnarray}

On using Eq.~(\ref{DERIV-9}) we see that the sum of the right-hand
side of Eq.~(\ref{DERIV-7}) is exactly equal to a certain $6$-$j$
symbol of Wigner. The $6$-$j$ symbols are scalar quantities which
arise in the construction of invariants from the vector-coupling
coefficients involving six angular momenta \cite{EDMONDS}. A
general $6$-$j$ symbols is written as
\begin{equation} \label{DEF-6j}
 \left\{\begin{array}{ccc}
 j_1 & j_2 & j_3 \\
 j_4 & j_5 & j_6
 \end{array}\right\}.
\end{equation}
For the sum of Eq.~(\ref{DERIV-7}) we have $j_1=j_2=j_4=j_5=j$,
$j_3=J$ and $j_6=K$. Hence, we finally obtain:
\begin{equation} \label{THETA-6j}
 \Theta_{JK} = \sqrt{(2J+1)(2K+1)}
 \left\{\begin{array}{ccc}
 j & j & J \\
 j & j & K
 \end{array}\right\}.
\end{equation}
This equation represents a central result of this paper. It yields
a general expression for the $\vartheta_2$-transformation in terms
of Wigner's $6$-$j$ symbols on which our investigation of the
structure of rotationally invariant states is based.

The $6$-$j$ symbols (\ref{DEF-6j}) are known to be invariant under
any permutation of their columns and under the interchange of the
upper and lower entries in any two columns. It follows that the
expression on the right-hand side of (\ref{THETA-6j}) is symmetric
with respect to the interchange of $J$ and $K$. It is also known
from the theory of angular momentum that the expression on the
right-hand side of (\ref{THETA-6j}) represents an orthogonal
matrix, in accordance with our previous considerations.

The properties of the $6$-$j$ symbols have been studied in great
detail and many explicit expressions and closed formulae are
known. Computational methods and recursion relations for the
$6$-$j$ symbols may be found in \cite{EDMONDS}. Equation
(\ref{THETA-6j}) enables one to employ these results in the
determination of the matrix $\Theta$. For example, the first two
rows and columns of $\Theta$ are given by
\begin{eqnarray}
 \Theta_{J0} &=& \Theta_{0J} = \frac{\sqrt{2J+1}}{N} (-1)^{2j+J},
 \label{THETA-J0} \\
 \Theta_{J1} &=& \Theta_{1J} = \sqrt{3(2J+1)}
 \frac{(N-1)(N+1)-2J(J+1)}{N(N-1)(N+1)} \nonumber \\
 && \qquad \qquad \times (-1)^{2j+1+J}. \label{THETA-J1}
\end{eqnarray}

Being real symmetric and orthogonal, the matrix $\Theta$ can of
course be diagonalized and has eigenvalues $\pm 1$. The
eigenvectors of $\Theta$ may be found from the sum rules for the
$6$-$j$ symbols given in \cite{EDMONDS}. If we write the sum rule
involving products of two $6$-$j$ symbols in terms of the matrix
elements $\Theta_{JK}$ we get
\begin{equation}
 \sum_K \Theta_{JK} (-1)^K \Theta_{KL}
 = (-1)^{L} (-1)^J \Theta_{JL}.
\end{equation}
We infer from this equation that the vector ${\bm{\alpha}}^{(L)}$
with components $\alpha_J^{(L)}=(-1)^{J}\Theta_{JL}$ is an
eigenvector of $\Theta$ with eigenvalue $(-1)^{L}$. Once we have
determined the matrix $\Theta$ we can therefore immediately write
its eigenvectors: One multiplies for all $J$ the $J$-th row of
$\Theta$ by $(-1)^J$; the columns of the resulting matrix then
represent the eigenvectors of $\Theta$.

It follows from the orthogonality of the matrix $\Theta$ that the
vectors ${\bm{\alpha}}^{(L)}$, $L=0,1,\ldots,2j$, form an
orthonormal basis of the parameter space. After a transformation
to principal axes $\Theta$ therefore takes the form
${\mbox{diag}}(+1,-1,+1,\ldots,(-1)^{2j})$, which describes a
reflection of the principal axes belonging to the eigenvalue $-1$.
The trace of $\Theta$ is obviously equal to zero for $N$ even
(half-integer $j$), and equal to $1$ for $N$ odd (integer $j$).

According to Eq.~(\ref{THETA-J0}) the components of the first
eigenvector ${\bm{\alpha}}^{(0)}$ are given by
$\alpha^{(0)}_J=(-1)^J\Theta_{J0}=(-1)^{2j}\sqrt{2J+1}/N$. This
vector is proportional to the vector which represents the state of
maximal entropy. The first eigenvector equation
$\Theta{\bm{\alpha}}^{(0)}={\bm{\alpha}}^{(0)}$ thus expresses the
invariance of the state of maximal entropy under $\vartheta_2$. It
may be written as
\begin{equation}
 \sum_{K=0}^{2j} \Theta_{JK} \frac{\sqrt{2K+1}}{N}
 = \frac{\sqrt{2J+1}}{N}.
\end{equation}
This equation can also be used to check that $\Theta$ preserves
the normalization (\ref{NORMIERUNG}).

\section{$SO(3)$-invariant PPT states}\label{PPT-STATES}

\subsection{Geometric representation}
We define $S_p$ to be the set of $SO(3)$-invariant PPT states,
i.~e., the set of rotationally invariant states which are positive
under $\vartheta_2$ (or, equivalently, under $T_2$). The
properties of $\vartheta_2$ imply that $S_p$ is the set of density
matrices $\rho$ for which $\rho'=\vartheta_2\rho$ is again a
density matrix and that $S_p$ is the intersection of $S$ with its
image under $\vartheta_2$:
\begin{equation} \label{D-P}
 S_p = S \cap \vartheta_2 S.
\end{equation}
Since $S$ is an $(N-1)$-simplex and $\vartheta_2$ is a
non-singular transformation the set $\vartheta_2 S$ is again an
$(N-1)$-simplex. Being the intersection of two convex sets, $S_p$
is also a convex set.

With the help of the properties of the matrix $\Theta$ derived in
Sec. \ref{PROP-THETA} it is easy to give the geometric
construction of $S_p$ employing the space of the
$\alpha_J$-parameters: One takes the $(N-1)$-simplex describing
$S$ and determines the intersection with its image under the
linear map given by the matrix $\Theta$. Since $S$ is convex it
suffices to determine the images of the extreme points of $S$ in
order to construct $\vartheta_2 S$.

To facilitate the geometric visualization we shall use in the
following an $(N-1)$-dimensional parameter space: A Hermitian and
rotationally invariant operator of trace $1$ is characterized
uniquely by $(N-1)$ real parameters $(\alpha_0,
\alpha_2,\ldots,\alpha_{2j-1})$. This means that we eliminate the
parameter $\alpha_{2j}$ by means of Eq.~(\ref{NORMIERUNG}) which
expresses the condition of unit trace. The state space $S$ can
then be identified with an $(N-1)$-simplex in ${\mathbb R}^{N-1}$
which is given by the conditions:
\begin{equation} \label{D-SPACE}
 \sum_{J=0}^{2j-1} \frac{\sqrt{2J+1}}{N} \alpha_J
 \leq 1, \qquad \alpha_0,\ldots,\alpha_{2j-1} \geq 0.
\end{equation}

\subsection{Examples}\label{EXAMPLES-1}
We illustrate the geometric construction of the set $S_p$ of PPT
states for $N=2$, $3$ and $4$. It will be seen that $S_p$ is
isomorphic to an $(N-1)$-dimensional cube. The matrix elements
$\Theta_{JK}$ can by determined with the help of
Eqs.~(\ref{THETA-J0}) and (\ref{THETA-J1}) and by use of the
general properties of $\Theta$ described in Sec.~\ref{PROP-THETA}.

\subsubsection{$2\otimes 2$ systems}\label{2times2}
In the simplest case $N=2$ the total system consists of two
particles with spin $j=\frac{1}{2}$ (two qubits). The total
angular momentum thus takes the values $J=0,1$ such that we can
use a single parameter $\alpha_0$ to describe a rotationally
invariant Hermitian operator of unit trace. The inequalities
(\ref{D-SPACE}) yield $0 \leq \alpha_0 \leq 2$. The space of
rotationally invariant density matrices is therefore given by the
interval ($1$-simplex) $S=[0,2]$. The matrix $\Theta$ is found to
be
\begin{equation}
 \Theta = \frac{1}{2} \left( \begin{array}{cc}
 -1 & \sqrt{3} \\
 \sqrt{3} & 1
 \end{array} \right),
\end{equation}
which is obviously symmetric, orthogonal and of trace zero. The
condition (\ref{NORMIERUNG}) gives
$\alpha_1=\frac{1}{\sqrt{3}}(2-\alpha_0)$ which is used to
eliminate $\alpha_1$ from the transformation
$\bm{\alpha}'=\Theta\bm{\alpha}$. One finds that $\vartheta_2$
maps the point $\alpha_0=0$ to $\alpha'_0=1$ and the point
$\alpha_0=2$ to $\alpha'_0=-1$. This yields
$\vartheta_2S=[-1,+1]$, and, hence, we get the set of PPT states:
\begin{equation}
 S_p = S\cap\vartheta_2S = [0,1].
\end{equation}

We note that for the present case of two dimensions the rotational
invariance is equivalent to the invariance under all product
unitaries $U\otimes U$. The states constructed above are therefore
identical to the Werner states of $2\otimes 2$ systems.

\subsubsection{$3\otimes 3$ systems}\label{3times3}
For $N=3$ (two qutrits) we have $j=1$ and $J=0,1,2$. We can
therefore use two parameters $(\alpha_0,\alpha_1)$ to characterize
a Hermitian and rotationally invariant operator with trace $1$.
Equation (\ref{D-SPACE}) now yields that the set $S$ of invariant
states is given by the inequalities:
\begin{equation} \label{D-SPACE-N3}
 \frac{1}{3} \alpha_0 +  \frac{1}{\sqrt{3}} \alpha_1
 \leq 1, \qquad \alpha_0, \alpha_1 \geq 0.
\end{equation}
Hence, $S$ is a triangle (2-simplex) with vertices $A=(0,0)$,
$B=(3,0)$ and $C=(0,\sqrt{3})$.

\begin{figure}[htb]
\includegraphics[width=\linewidth]{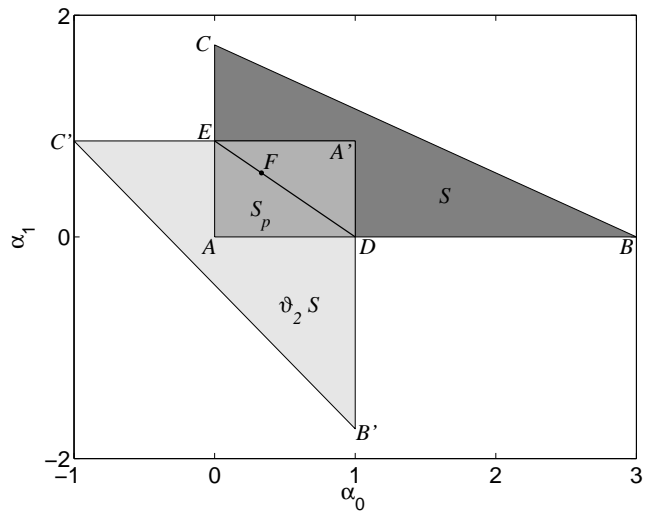}
\caption{$SO(3)$-invariant Hermitian operators of trace $1$ for
$3\otimes 3$ systems. Triangle $ABC$: The set $S$ of rotationally
invariant density matrices. Triangle $A'B'C'$: The transform
$\vartheta_2S$. Rectangle $ADA'E$: The set $S_p$ of PPT states,
which is equal to the set $S_s$ of separable states (see
Sec.~\ref{3times3-sep}). The line $DE$ represents the fixed points
of $\vartheta_2$, the point $F$ the state of maximal entropy and
$B$ the singlet state $|00\rangle$. \label{fig1}}
\end{figure}

The matrix $\Theta$ now becomes:
\begin{equation}
 \Theta = \frac{1}{3} \left( \begin{array}{ccc}
 1 & -\sqrt{3} & \sqrt{5} \\
 -\sqrt{3} & \frac{3}{2} & \frac{\sqrt{15}}{2} \\
 \sqrt{5} & \frac{\sqrt{15}}{2} & \frac{1}{2}
 \end{array} \right).
\end{equation}
One easily verifies that this is a symmetric and orthogonal matrix
of trace $1$. On eliminating the parameter $\alpha_{2}$ we find
that $\vartheta_2$ acts as follows on the vertices of $S$:
\begin{eqnarray}
 A=\left(0,0\right) &\mapsto& A'=\left(1,\frac{\sqrt{3}}{2}\right), \\
 B=\left(3,0\right) &\mapsto& B'=(1,-\sqrt{3}), \\
 C=(0,\sqrt{3}) &\mapsto& C'=\left(-1,\frac{\sqrt{3}}{2}\right).
\end{eqnarray}
Thus, $\vartheta_2S$ is the triangle with vertices $A'$, $B'$ and
$C'$.

The sets $S$ and $\vartheta_2S$ are depicted in Fig.~\ref{fig1}.
The figure also shows the line of the fixed points of
$\vartheta_2$ with endpoints $D=(1,0)$ and
$E=\left(0,\frac{\sqrt{3}}{2}\right)$. This line is easily
determined from the matrix $\Theta$ and its eigenvectors. Being
invariant under $\vartheta_2$, the point $F$, which describes the
state of maximal entropy, lies of course on this line.

The rectangle with vertices $A$, $D$, $A'$ and $E$ represents the
intersection $S_p=S\cap\vartheta_2S$ of the PPT states. It should
be noted that the rotational invariance in the present example is
equivalent to the invariance under the product transformations
$O\otimes O$, where $O$ varies over the group of orthogonal
$3\times 3$ matrices \cite{VOLLBRECHT}.

\subsubsection{$4\otimes 4$ systems}\label{4times4}
The case $N=4$ corresponds to a system composed of two particles
with spin $j=\frac{3}{2}$. The total angular momentum assumes the
values $J=0,1,2,3$. Thus we get a three-dimensional parameter
space with parameters $(\alpha_0,\alpha_1,\alpha_2)$. By virtue of
Eq.~(\ref{D-SPACE}) the set of rotationally invariant states is
determined by the inequalities:
\begin{equation} \label{D-SPACE-N4}
 \frac{1}{4} \alpha_0 +  \frac{\sqrt{3}}{4} \alpha_1
 + \frac{\sqrt{5}}{4} \alpha_2
 \leq 1, \qquad \alpha_0, \alpha_1, \alpha_2 \geq 0.
\end{equation}
This shows that $S$ is a tetrahedron (3-simplex) with vertices
$A=(0,0,0)$, $B=\left(4,0,0\right)$,
$C=\left(0,\frac{4}{\sqrt{3}},0\right)$ and
$D=\left(0,0,\frac{4}{\sqrt{5}}\right)$.

\begin{figure}[htb]
\includegraphics[width=\linewidth]{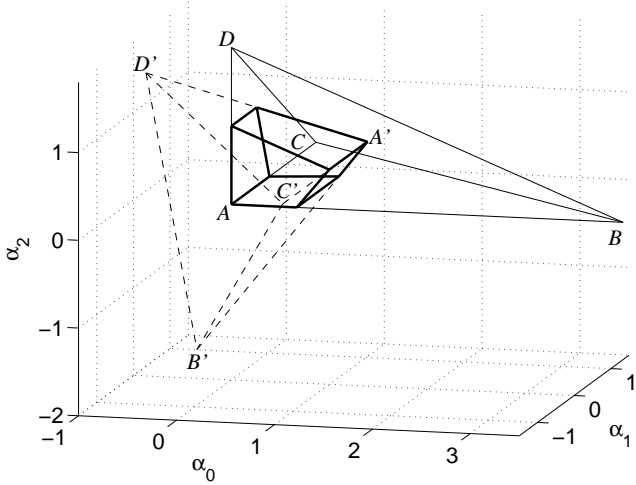}
\caption{$SO(3)$-invariant Hermitian operators of trace $1$ for
$4\otimes 4$ systems. The tetrahedron $ABCD$ (continuous lines)
represents the set $S$ of invariant states, and the tetrahedron
$A'B'C'D'$ (broken lines) its transform $\vartheta_2S$. The
intersection (bold lines) is the set $S_p$ of the PPT states.
\label{fig2}}
\end{figure}

The matrix $\Theta$ is given by:
\begin{equation}
 \Theta = \frac{1}{4} \left( \begin{array}{cccc}
 -1 & \sqrt{3} & -\sqrt{5} & \sqrt{7} \\
 \sqrt{3} & -\frac{11}{5} & \sqrt{\frac{3}{5}} & \frac{3\sqrt{21}}{5} \\
 -\sqrt{5} & \sqrt{\frac{3}{5}} & 3 & \sqrt{\frac{7}{5}} \\
 \sqrt{7} & \frac{3\sqrt{21}}{5} & \sqrt{\frac{7}{5}} & \frac{1}{5}
 \end{array} \right).
\end{equation}
One checks that this matrix is symmetric, orthogonal and of trace
zero. It leads to the following mapping of the vertices of the
tetrahedron $S$ under $\vartheta_2$:
\begin{eqnarray}
 A=\left(0,0,0\right) &\mapsto&
 A'=\left(1,\frac{3\sqrt{3}}{5},\frac{1}{\sqrt{5}}\right), \label{AA'} \\
 B=\left(4,0,0\right) &\mapsto&
 B'=\left(-1,\sqrt{3},-\sqrt{5}\right), \\
 C=\left(0,\frac{4}{\sqrt{3}},0\right) &\mapsto&
 C'=\left(1,-\frac{11}{5\sqrt{3}},\frac{1}{\sqrt{5}}\right), \\
 D=\left(0,0,\frac{4}{\sqrt{5}}\right) &\mapsto&
 D'=\left(-1,\frac{\sqrt{3}}{5},\frac{3}{\sqrt{5}}\right).
\end{eqnarray}
The points $A'$, $B'$, $C'$ and $D'$ are the vertices of the
transformed tetrahedron $\vartheta_2S$, as shown in
Fig.~\ref{fig2}.

\begin{figure}[htb]
\includegraphics[width=\linewidth]{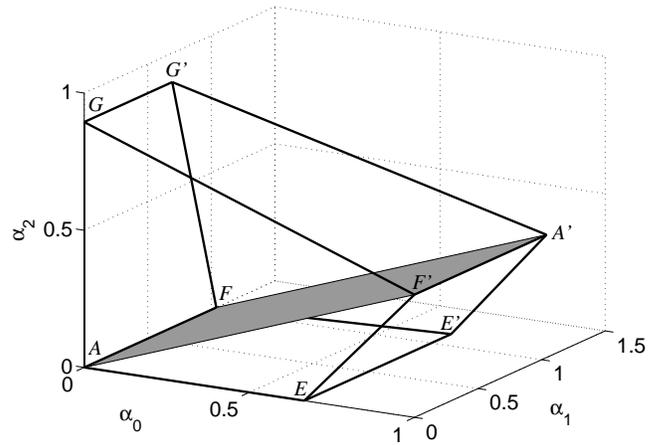}
\caption{Enlarged picture of the cube $S_p$ of PPT states (see
Fig.~\ref{fig2}). The plane $AA'FF'$ subdivides $S_p$ into two
prisms. The prism $AA'FF'GG'$ represents the set $S_s$ of
separable states (see Sec.~\ref{4times4-sep}). \label{fig3}}
\end{figure}

We see from Fig.~\ref{fig2} that the intersection $S_p=S\cap
\vartheta_2S$ is isomorphic to a 3-dimensional cube. An enlarged
picture of this cube is shown in Fig.~\ref{fig3}. The vertices of
$S_p$ are given by the points $A$, $A'$ and
\begin{eqnarray}
 E=\left(\frac{2}{3},0,0\right), &&
 E'=\left(\frac{2}{3},\frac{2\sqrt{3}}{3},0\right), \label{EE'} \\
 F=\left(0,\frac{3\sqrt{3}}{5},0\right), &&
 F'=\left(1,0,\frac{1}{\sqrt{5}}\right), \label{FF'} \\
 G=\left(0,0,\frac{2}{\sqrt{5}}\right), &&
 G'=\left(0,\frac{2\sqrt{3}}{5},\frac{2}{\sqrt{5}}\right). \label{GG'}
\end{eqnarray}
These points may be obtained as follows (see Fig.~\ref{fig3}). One
takes the three edges emerging from the vertex $A'$ of the
tetrahedron $\vartheta_2 S$ and determines their intersection with
the faces of the tetrahedron $S$. This yields the points $E'$,
$F'$ and $G'$. The points $E$, $F$ and $G$ are then given by the
images of $E'$, $F'$ and $G'$ under $\vartheta_2$.

\section{Separable states}\label{SEP}

\subsection{Construction of $SO(3)$-invariant separable
            states}\label{SEP-STATES}
The set of separable states is defined to be the set of states
$\rho$ which can be written as a convex sum of product states:
\begin{equation} \label{DEF-SEP}
 \rho = \sum_i \lambda_i \rho_i^{(1)} \otimes \rho_i^{(2)},
 \qquad \lambda_i \geq 0, \qquad \sum_i \lambda_i = 1,
\end{equation}
where the $\rho_i^{(1)}$ and $\rho_i^{(2)}$ are normalized local
states \cite{WERNER}. It follows from this definition and from the
positivity of $\vartheta$ that the map $\vartheta_2$ is positive
on separable states. Thus, $\vartheta_2$ maps rotationally
invariant and separable states to rotationally invariant and
separable states.

We denote the set of $SO(3)$-invariant separable states by $S_s$.
This set is contained in the set of states which are positive
under $\vartheta_2$:
\begin{equation} \label{D-S}
 S_s \subset S_p
 = S \cap \vartheta_2 S.
\end{equation}
This equation expresses the Peres PPT criterion. It can easily be
applied in the present formulation once the matrix $\Theta$ has
been determined: Given a rotationally invariant state $\rho$ in
terms of its parameter vector $\bm{\alpha}$ according to
Eq.~(\ref{D-REP}), a necessary condition for this state to be
separable is that all components of the transformed parameter
vector ${\bm{\alpha}}'=\Theta {\bm{\alpha}}$ are positive.

To fully characterize the set of separable states one introduces a
projection super-operator $\Pi$, also known as twirl operator.
Given any state $\rho$ of the bipartite system the operator
\begin{equation} \label{PROJECTION}
 \Pi\rho = \sum_{J=0}^{2j} \frac{1}{2J+1} P_{J} {\mbox{tr}}\{P_J\rho\}
\end{equation}
is positive, of trace $1$ and rotationally invariant. The map
$\rho \mapsto \Pi\rho$ defines a projection (i.~e., $\Pi^2=\Pi$)
from the total state space onto the space $S$ of rotationally
invariant states. Moreover, if $\rho$ is separable then $\Pi\rho$
is again separable.

We see from Eq.~(\ref{PROJECTION}) that the $\alpha_J$-parameters
corresponding to the projection $\Pi\rho$ are given by
$\alpha_J=\frac{N}{\sqrt{2J+1}}{\mathrm{tr}}\{P_J\rho\}$. If we
take a pure product state
\begin{equation} \label{PRODUCT}
 \rho =
 |\varphi^{(1)}\varphi^{(2)}\rangle\langle\varphi^{(1)}\varphi^{(2)}|
\end{equation}
involving normalized local states $|\varphi^{(1)}\rangle$ and
$|\varphi^{(2)}\rangle$, the $\alpha_J$-parameters of its
projection are found to be
\begin{equation} \label{FUNC-1}
 \alpha_J = \tilde{\alpha}_J[\varphi^{(1)},\varphi^{(2)}]
 = \frac{N}{\sqrt{2J+1}}
 \langle\varphi^{(1)}\varphi^{(2)}|P_J|\varphi^{(1)}\varphi^{(2)}\rangle.
\end{equation}

It is known that any separable state can be written as a convex
sum of pure product states. We define $W$ to be the range of the
parameter vector $\bm{\alpha}$ whose components $\alpha_J$ are
given by the above functionals
$\tilde{\alpha}_J[\varphi^{(1)},\varphi^{(2)}]$, where
$|\varphi^{(1)}\rangle$ and $|\varphi^{(2)}\rangle$ run
independently over all normalized states in ${\mathcal{H}}$. With
this definition one has the following result \cite{VOLLBRECHT}:
The set $S_s$ of rotationally invariant and separable states is
equal to the convex hull of the range $W$, i.~e., to the smallest
convex set containing $W$. Thus, we have:
\begin{equation}
 S_s = {\mbox{hull}}(W) \subset S_p.
\end{equation}
The determination of $S_s$ therefore amounts to the determination
of the convex hull of the range of the functionals
$\tilde{\alpha}_J[\varphi^{(1)},\varphi^{(2)}]$ given by
Eq.~(\ref{FUNC-1}). This task can be simplified by the following
observations.

First, since $S_s$ is the convex hull of $W$ which, in turn, is
contained in $S_p$, a good starting point is to consider the
extreme points (vertices) of $S_p$. If one finds, for example,
that all extreme points of $S_p$ belong to $W$ one concludes
immediately that $S_s$ must be identical to $S_p$.

Second, it is clear by construction that the functionals
$\tilde{\alpha}_J$ are invariant under simultaneous rotations
$|\varphi^{(1,2)}\rangle \mapsto D(R)|\varphi^{(1,2)}\rangle$ of
the input arguments. Pairs of state vectors differing by such a
transformation are thus projected to one and the same point of the
parameter space and need not be considered separately.

Third, the range $W$ is invariant under the map $\vartheta_2$.
This means that if the point $\bm{\alpha}$ belongs to $W$, then
also the transformed point $\Theta\bm{\alpha}$ belongs to $W$.
This statement can easily be proven by use of the results of
Sec.~\ref{TIME-REVERSAL}. In fact, we have
\begin{eqnarray} \label{FUNC-TRANSFORM}
 \tilde{\alpha}_J[\varphi^{(1)},\tau\varphi^{(2)}]
 &=& \frac{N}{\sqrt{2J+1}}
 \langle\varphi^{(1)}\varphi^{(2)}|\vartheta_2P_J|\varphi^{(1)}\varphi^{(2)}\rangle
 \nonumber \\
 &=& \sum_{K=0}^{2j} \Theta_{JK}
 \tilde{\alpha}_K[\varphi^{(1)},\varphi^{(2)}].
\end{eqnarray}
We see that the transformation
$\bm{\alpha}\mapsto\Theta\bm{\alpha}$ corresponds to the time
reversal transformation $\tau$ carried out on the second input
argument of the functionals. Equation (\ref{FUNC-TRANSFORM}) also
demonstrates that if $|\varphi^{(2)}\rangle$ is invariant under
$\tau$ the corresponding parameter vector represents, for any
choice of $|\varphi^{(1)}\rangle$, a fixed point of $\Theta$.

\subsection{Representation in terms of spherical tensors}
In addition to the projections $P_J$ there exist further
rotationally invariant operators which span the set $S$ and which
lead to a particularly useful representation of the set of
separable states. To construct these operators we introduce the
irreducible spherical tensor operators $T_{JM}$ acting on
${\mathcal{H}}$, where, as before, $J=0,1,\ldots,2j$ and
$M=-J,-J+1,\ldots,+J$. The matrix elements of these operators are
defined by the $3$-$j$ symbols:
\begin{equation} \label{DEF-TJM}
 \langle jm|T_{JM}|jm'\rangle = (-1)^{j-m} \sqrt{2J+1}
 \left(\begin{array}{ccc}
 j & j & J \\
 m & -m' & -M
 \end{array}\right).
\end{equation}
According to the selection rules of the $3$-$j$ symbols the matrix
element (\ref{DEF-TJM}) is zero for $\Delta m \equiv m-m' \neq M$.
The tensor operators $T_{JM}$ represent a complete system of
operators on ${\mathcal{H}}$ which are orthonormal with respect to
the Hilbert-Schmidt inner product, i.~e., one has
${\mathrm{tr}}\{T^{\dagger}_{JM}T_{J'M'}\}=\delta_{JJ'}\delta_{MM'}$.

For a fixed $J$ the $(2J+1)$ operators $T_{JM}$ transform
according to an irreducible representation of the rotation group
corresponding to the angular momentum $J$. For example, the
$T_{1M}$ transform as the spherical components of a vector, while
the $T_{2M}$ behave as the components of a second-rank tensor
under rotations. For $N=2$ the tensor components $T_{1M}$ may be
expressed in terms of the Pauli matrices as
$T_{10}=\frac{1}{\sqrt{2}}\sigma_3$ and $T_{1,\pm
1}=\mp\frac{1}{2}(\sigma_1 \pm i\sigma_2)$. The definition
(\ref{DEF-TJM}) leads to the relation
$T^{\dagger}_{JM}=T^T_{JM}=(-1)^MT_{J,-M}$. One concludes that the
tensor operators are eigen-operators of the time reversal
transformation: $\vartheta T_{JM} = (-1)^J T_{JM}$.

It follows from the transformation behaviour of the $T_{JM}$ that
the operators on the product space defined by
\begin{equation} \label{DEF-Q}
 Q_J = \sum_{M=-J}^{+J} T_{JM} \otimes T_{JM}^{\dagger}
\end{equation}
are invariant under rotations. The connection between the
projections $P_J$ and the operators $Q_J$ is provided by the
relation
\begin{equation} \label{PJ-QJ}
 \vartheta_2 P_J = Q_J {\mathbb F},
\end{equation}
where we have introduced the flip operator ${\mathbb F}$ which is
defined by
\begin{equation} \label{FLIP}
 {\mathbb F}|jm_1jm_2\rangle = |jm_2jm_1\rangle.
\end{equation}
The proof of Eq.~(\ref{PJ-QJ}) is given in Appendix \ref{APP-1}.

Equation (\ref{PJ-QJ}) leads to an alternative characterization of
the set of separable states. Since $|\varphi^{(1)}\rangle$ and
$|\varphi^{(2)}\rangle$ vary independently over all normalized
states we may use the right-hand side of
Eq.~(\ref{FUNC-TRANSFORM}) instead of the original expression
(\ref{FUNC-1}) for the functionals
$\tilde{\alpha}_J[\varphi^{(1)},\varphi^{(2)}]$. If we introduce
(\ref{PJ-QJ}) into (\ref{FUNC-TRANSFORM}) we find that we can
employ the functionals
\begin{equation} \label{FUNC-2}
 \tilde{\alpha}_J[\varphi^{(1)},\varphi^{(2)}]
 = \frac{N}{\sqrt{2J+1}} \sum_{M=-J}^{+J}
 |\langle\varphi^{(1)}|T_{JM}|\varphi^{(2)}\rangle|^2
\end{equation}
in order to construct the range $W$ and the set $S_s$ of separable
states. An advantage of this formulation is that it leads to a
very simple expression for $J=0$. Namely, since
$T_{00}=\frac{1}{\sqrt{N}}I$ we have
\begin{equation} \label{ALPHA-0}
 \tilde{\alpha}_0[\varphi^{(1)},\varphi^{(2)}]
 = |\langle\varphi^{(1)}|\varphi^{(2)}\rangle|^2.
\end{equation}

It might be interesting to note that Eq.~(\ref{PJ-QJ}) can be used
to identify the one-parameter family of the Werner states given by
\begin{equation}
 \rho_W = \frac{1}{N^3-N}\left[ (N-\lambda)I\otimes I
 + (N\lambda -1) {\mathbb F} \right],
\end{equation}
where $-1\leq \lambda \leq +1$. These states are invariant under
all product unitaries $U\otimes U$. Therefore, all states of the
family are, in particular, invariant under rotations and belong to
$S$. The parameters $\alpha^W_J$ corresponding to $\rho_W$ are
found to be
\begin{eqnarray}
 \alpha^W_J
 &=& \frac{N}{\sqrt{2J+1}}{\mathrm{tr}}\{P_J\rho_W\}
 \label{ALPHA-WERNER} \\
 &=& \frac{\sqrt{2J+1}}{N^2-1}
 \left[N-\lambda+(-1)^{2j+J}(N\lambda-1)\right]. \nonumber
\end{eqnarray}
To obtain this result one has to determine the expression
${\mathrm{tr}}\{P_J{\mathbb F}\}$. This may be done by noting that
for $J=0$ Eq.~(\ref{PJ-QJ}) yields ${\mathbb F}=N\vartheta_2P_0$.
The expression ${\mathrm{tr}}\{P_J{\mathbb F}\}$ can therefore be
written in terms of the matrix elements $\Theta_{J0}$ which are
given by Eq.~(\ref{THETA-J0}). The family of the isotropic states
can be embedded in a similar way into $S$ if one first performs
the local unitary transformation $I\otimes V$.

\subsection{Examples}\label{EXAMPLES-2}
We construct the set $S_s$ of separable states for the examples
considered in Sec.~\ref{EXAMPLES-1}. To this end, we make use of
the functionals (\ref{FUNC-2}) which characterize $S_s$ and of the
general properties of the range $W$ described in
Sec.~\ref{SEP-STATES}.

\subsubsection{$2\otimes 2$ systems}\label{2times2-sep}
In the case of our first example discussed in Sec.~\ref{2times2}
we found that the parameter $\alpha_0$ describes a PPT state if
and only if  $\alpha_0\in S_p = [0,1]$. We immediately see from
Eq.~(\ref{ALPHA-0}) that the set of separable states and the set
of PPT states are identical, that is $S_s=S_p$. In fact, according
to Eq.~(\ref{ALPHA-0}) the functional
$\tilde{\alpha}_0[\varphi^{(1)},\varphi^{(2)}]$ can take any value
in the interval $[0,1]$ because $|\varphi^{(1,2)}\rangle$ are
arbitrary normalized states. This shows that in the present case
positivity under $\vartheta_2$ is a necessary and sufficient
condition for separability, which is a well-known fact
\cite{HORODECKI96b}.

\subsubsection{$3\otimes 3$ systems}\label{3times3-sep}
Using the results of Sec.~\ref{3times3} we show that also for
$3\otimes 3$ systems the set of PPT states and the set of
separable states coincide, i.~e., $S_s=S_p$. Thus, positivity
under $\vartheta_2$ is again a necessary and sufficient condition
for separability in this case, as has been demonstrated by
Vollbrecht and Werner \cite{VOLLBRECHT}. To prove this we verify
that the extreme points of $S_p$, that is, the points $A$, $A'$,
$D$ and $E$ belong to the range $W$ (see Fig.~\ref{fig1}).

First, we choose $|\varphi^{(1)}\rangle=|j=1, m=+1\rangle$ and
$|\varphi^{(2)}\rangle=|j=1,m=-1\rangle$. These states are
orthogonal and, hence, $\tilde{\alpha}_0=0$ according to
Eq.~(\ref{ALPHA-0}). Using the selection rules for the matrix
elements (\ref{DEF-TJM}) of the tensor operators $T_{1M}$ one sees
that also $\tilde{\alpha}_1=0$ (the operators $T_{1M}$ cannot
connect states whose magnetic quantum numbers differ by $2$). This
shows that $A=(0,0)$ belongs to the range $W$. It also follows
that $A'$ belongs to $W$, because $A'$ is the image of $A$ under
$\vartheta_2$.

Next, we consider the state
\begin{equation}
 |\varphi^{(2)}\rangle=\frac{1}{\sqrt{2}}(|1,+1\rangle+|1,-1\rangle).
\end{equation}
This state is invariant under the time reversal transformation
(\ref{TAU}). Thus, for any choice of $|\varphi^{(1)}\rangle$, the
point $(\tilde{\alpha}_0,\tilde{\alpha}_1)$ is a fixed point of
$\vartheta_2$ and, hence, belongs to the line $DE$ (see
Fig.~\ref{fig1}). If $|\varphi^{(1)}\rangle$ is any state
orthogonal to $|\varphi^{(2)}\rangle$ we have that, additionally,
$\tilde{\alpha}_0=0$ and, hence,
$(\tilde{\alpha}_0,\tilde{\alpha}_1)=\left(0,\frac{\sqrt{3}}{2}\right)\equiv
E\in W$. One the other hand, if we take
$|\varphi^{(1)}\rangle=|\varphi^{(2)}\rangle$, then
$\tilde{\alpha}_0=1$ and, therefore,
$(\tilde{\alpha}_0,\tilde{\alpha}_1)=(1,0)\equiv D\in W$. This
concludes the proof.

\subsubsection{$4\otimes 4$ systems}\label{4times4-sep}
For $4\otimes 4$ systems it is again possible to give a complete
geometric construction of the set of separable states by use of
the results of Sec.~\ref{4times4}. In the case $N=4$ we have to
consider the following functionals:
\begin{eqnarray}
 \tilde{\alpha}_0[\varphi^{(1)},\varphi^{(2)}]
 \! &=& \! |\langle\varphi^{(1)}|\varphi^{(2)}\rangle|^2,
 \label{ALPHA-00} \\
 \tilde{\alpha}_1[\varphi^{(1)},\varphi^{(2)}]
 \! &=& \! \frac{4}{\sqrt{3}} \sum_{M=-1}^{+1}
 |\langle\varphi^{(1)}|T_{1M}|\varphi^{(2)}\rangle|^2,
 \label{ALPHA-1} \\
 \tilde{\alpha}_2[\varphi^{(1)},\varphi^{(2)}]
 \! &=& \! \frac{4}{\sqrt{5}} \sum_{M=-2}^{+2}
 |\langle\varphi^{(1)}|T_{2M}|\varphi^{(2)}\rangle|^2.
 \label{ALPHA-2}
\end{eqnarray}
To construct $S_s$ we proceed in four steps, investigating the
extreme points of $S_p$ given by Eqs.~(\ref{AA'}), (\ref{EE'}),
(\ref{FF'}) and (\ref{GG'}) (see Fig.~\ref{fig3}).

(1) We show that $A,A'\in W$. To proof this we take
$|\varphi^{(1)}\rangle=|\frac{3}{2},+\frac{3}{2}\rangle$ and
$|\varphi^{(2)}\rangle=|\frac{3}{2},-\frac{3}{2}\rangle$. These
states are orthogonal and, therefore, $\tilde{\alpha}_0=0$
according to Eq.~(\ref{ALPHA-00}). Since the magnetic quantum
numbers of the states differ by $3$ the selection rules for the
matrix elements (\ref{DEF-TJM}) yield that $\langle
\varphi^{(1)}|T_{JM}|\varphi^{(2)}\rangle=0$ for $J=1,2$. Thus,
Eqs.~(\ref{ALPHA-1}) and (\ref{ALPHA-2}) yield
$\tilde{\alpha}_1=\tilde{\alpha}_2=0$. Hence, $A=(0,0,0)$ and
$A'=\vartheta_2 A$ belong to $W$.

(2) We demonstrate that also $G,G'\in W$. To this end, we take
$|\varphi^{(1)}\rangle=|\frac{3}{2},+\frac{3}{2}\rangle$ and
$|\varphi^{(2)}\rangle=|\frac{3}{2},-\frac{1}{2}\rangle$. These
states are again orthogonal and we get $\tilde{\alpha}_0=0$. Since
$\Delta m = 2$ the matrix elements $\langle
\varphi^{(1)}|T_{1M}|\varphi^{(2)}\rangle$ vanish and, therefore,
$\tilde{\alpha}_1=0$.  The only matrix element of the $T_{2M}$
which is not equal to zero on account of the selection rules is
given by
\begin{equation}
 \langle\varphi^{(1)}|T_{22}|\varphi^{(2)}\rangle
 = \frac{1}{\sqrt{2}}.
\end{equation}
Thus, with the help of Eq.~(\ref{ALPHA-2}) we obtain
$\tilde{\alpha}_2=\frac{2}{\sqrt{5}}$. One concludes that
$G=(0,0,\frac{2}{\sqrt{5}})$ and, hence, also $G'=\vartheta_2 G$
belong to $W$.

(3) We claim that $F,F'\in W$. To prove this we choose the states:
\begin{eqnarray}
 |\varphi^{(1)}\rangle &=& \frac{1}{\sqrt{2}} \left(
 \left|\frac{3}{2},+\frac{3}{2}\right\rangle
 + \left|\frac{3}{2},-\frac{3}{2}\right\rangle \right), \\
 |\varphi^{(2)}\rangle &=& \frac{1}{\sqrt{2}} \left(
 \left|\frac{3}{2},+\frac{3}{2}\right\rangle
 - \left|\frac{3}{2},-\frac{3}{2}\right\rangle \right).
\end{eqnarray}
These states are obviously orthogonal and we get again
$\tilde{\alpha}_0=0$. The selection rules now yield $\langle
\varphi^{(1)}|T_{JM}|\varphi^{(2)}\rangle=0$ for $J=1,2$ and
$M\neq 0$, while
\begin{eqnarray} \label{TJ0}
 \langle \varphi^{(1)}|T_{J0}|\varphi^{(2)}\rangle
 &=& \frac{1}{2} \left\langle\frac{3}{2},+\frac{3}{2}\right|T_{J0}
 \left|\frac{3}{2},+\frac{3}{2}\right\rangle \nonumber \\
 &-& \frac{1}{2} \left\langle\frac{3}{2},-\frac{3}{2}\right|T_{J0}
 \left|\frac{3}{2},-\frac{3}{2}\right\rangle.
\end{eqnarray}
We have the following general relation between the matrix elements
of the tensor operators:
\begin{equation} \label{SYMMETRY-TJM}
 \langle j,-m|T_{J0}|j,-m\rangle = (-1)^J
 \langle j,+m|T_{J0}|j,+m\rangle.
\end{equation}
On using this we see that the expression (\ref{TJ0}) vanishes for
$J=2$. It follows that $\tilde{\alpha}_2=0$. On the other hand,
for $J=1$ we obtain
\begin{equation}
 \langle \varphi^{(1)}|T_{10}|\varphi^{(2)}\rangle
 = \left\langle\frac{3}{2},+\frac{3}{2}\right|T_{10}
 \left|\frac{3}{2},+\frac{3}{2}\right\rangle = \frac{3}{2\sqrt{5}}.
\end{equation}
With the help of (\ref{ALPHA-1}) this leads to
$\tilde{\alpha}_1=\frac{3\sqrt{3}}{5}$. In summary, we see that
$F=(0,\frac{3\sqrt{3}}{5},0)$ and $F'=\vartheta_2 F$ belong to the
range $W$.

(4) It is shown in Appendix \ref{APP-2} that the functionals
(\ref{ALPHA-00}) and (\ref{ALPHA-2}) fulfill the inequality:
\begin{equation} \label{ALPHA-INEQ}
 \tilde{\alpha}_2[\varphi^{(1)},\varphi^{(2)}] \geq
 \frac{1}{\sqrt{5}}
 \tilde{\alpha}_0[\varphi^{(1)},\varphi^{(2)}].
\end{equation}
It follows that $E$ and $E'$ do {\textit{not}} belong to the range
$W$. Namely, for these points we must have $\tilde{\alpha}_2=0$
and $\tilde{\alpha}_0=\frac{2}{3}$ [see Eq.~(\ref{EE'})] which
contradicts the inequality (\ref{ALPHA-INEQ}). This shows that in
the present case $S_s$ is a true subset of $S_p$, i.~e.,
positivity under $\vartheta_2$ is a necessary but not sufficient
condition for separability.

According to Sec. \ref{4times4} the set $S_s$ of separable states
is contained in the cube $S_p$ of the PPT states (see
Fig.~\ref{fig3}). By the above results the points $A$, $A'$, $F$,
$F'$, $G$, and $G'$ are contained in the range $W$. Since $S_s$ is
the convex hull of $W$ we conclude that $S_s$ contains {\textit{at
least}} the polyhedron $AA'FF'GG'$. We observe that this
polyhedron is isomorphic to a prism.

The inequality (\ref{ALPHA-INEQ}) yields an additional condition
for the separable states. It implies that all points of the range
$W$ must lie on or above the plane which is defined by $\alpha_2 =
\frac{1}{\sqrt{5}}\alpha_0$ and which is indicated as gray surface
in Fig.~\ref{fig3}. We note that according to Eqs.~(\ref{AA'}) and
(\ref{FF'}) the points $A$, $A'$, $F$ and $F'$ belong to this
plane. It follows that the set $S_s$ of separable states is in
fact {\textit{identical}} to the prism $AA'FF'GG'$.

In summary, the convex structure of the set of $SO(3)$-invariant
states of $4\otimes 4$ systems may be described by the following
inclusions:
\begin{equation} \label{INCLUSIONS}
 (\mbox{prism}\; S_s) \subset (\mbox{cube}\; S_p) \subset
 (\mbox{tetrahedron}\; S).
\end{equation}
The tetrahedron $S$, representing the set of all invariant states,
decomposes into the cube $S_p$ of PPT states and the set
$S\setminus S_p$ of entangled states whose partial transposition
has negative eigenvalues. The cube $S_p$ of PPT states, in turn,
consists of the prism $S_s$ of separable states and of the set
$S_p\setminus S_s$ of entangled PPT states. The plane
$\alpha_2=\frac{1}{\sqrt{5}}\alpha_0$ thus separates the entangled
PPT states from the separable states.

As can be seen from Fig.~\ref{fig3} the set $S_p\setminus S_s$ is
isomorphic to a prism from which one face has been removed. All
states belonging to this set are inseparable and have positive
partial transposition. This leads to the important conclusion that
$S_p\setminus S_s$ represents a three-dimensional manifold of
bound entangled states, i.~e., states which cannot be distilled by
local quantum operations and classical communication
\cite{BENNETT,VINCENZO,HORODECKI98}.

\section{Discussion and conclusions}\label{CONCLU}
We have analyzed the structure of the state spaces of bipartite
$N\otimes N$ systems which are invariant under product
representations of the rotation group. The main tool of the
analysis is the positive map $\vartheta$ which is unitarily
equivalent to the transposition $T$ and describes the behaviour of
local states under time reversal. Employing the properties of
$\vartheta$ one relates the partial time reversal
$\vartheta_2=I\otimes \vartheta$ to a linear transformation of the
parameter space ${\mathbb R}^N =\{\bm{\alpha}\}$ and expresses the
corresponding matrix $\Theta$ in terms of Wigner's $6$-$j$
symbols. This matrix has been used to obtain geometrical
representations for the sets of the separable and of the PPT
states in the cases $N=2$, $3$ and $4$.

In Sec.~\ref{4times4-sep} the inequality (\ref{ALPHA-INEQ})
enabled the construction of the set of separable states. Taken
together with the Peres PPT criterion this inequality yields a
necessary and sufficient condition for the separability of
rotationally invariant states of $4\otimes 4$ systems. It is of
great interest to examine the possibility of an extension of this
picture to higher dimensions. In this context it is important to
observe that the inequality (\ref{ALPHA-INEQ}) expresses the
positivity of a certain map $\Phi$ which is given by
\begin{equation} \label{PHI-POS}
 \Phi B = \sum_{M=-2}^{+2} T_{2M}BT_{2M}^{\dagger}
 - T_{00} B T_{00}^{\dagger}.
\end{equation}
This map is non-decomposable and detects all entangled PPT states.
Hence, we need exactly two maps, namely $\vartheta$ and $\Phi$, in
order to identify uniquely all separable states. These maps yield
complementary conditions for separability in the sense that the
two inequalities
\begin{equation}
 (I\otimes\vartheta)\rho \geq 0 \qquad {\mbox{and}} \qquad
 (I\otimes\Phi)\rho \geq 0
\end{equation}
constitute a necessary and sufficient separability criterion. It
should also be noted that the proof of Appendix \ref{APP-2} does
not rely on any invariance requirement. We conclude that
positivity under the map $\Phi_2=I\otimes \Phi$ is a necessary
condition of separability for all (not necessarily rotationally
invariant) states of $4\otimes 4$ systems.

The positive map introduced in Eq.~(\ref{PHI-POS}) corresponds to
an entanglement witness \cite{HORODECKI96a,TERHAL,LEWENSTEIN00}
which is given by the operator ${\mathcal{W}}=P_2-P_0$. The plane
$\alpha_2=\frac{1}{\sqrt{5}}\alpha_0$ in parameter space may be
viewed as an optimal hyperplane defined by this witness
${\mathcal{W}}$. This fact leads to the following interpretation
of the inequality (\ref{ALPHA-INEQ}): If a measurement of the
total angular momentum is carried out on a separable state, the
probability of finding the value $J=2$ must be larger or equal to
the probability of finding the value $J=0$.

The method developed here suggests many generalizations and
applications. An obvious extension is to consider bipartite
systems whose local state spaces are not isomorphic, involving two
different angular momenta $j^{(1)} \neq j^{(2)}$. Further
important topics are an extension of the analysis given in
Sec.~\ref{EXAMPLES-2} to higher-dimensional systems, the treatment
of other symmetry groups, and entanglement in multipartite
systems.

The matrix $\Theta$ contains the complete information on the
behaviour of the spectrum of the invariant states under partial
transposition. It can also be used to express various separability
criteria and entanglement measures and to design positive maps and
entanglement witnesses. Examples of applications are the
determination of the relative entropy of entanglement with respect
to the set of PPT states \cite{AUDENAERT}, and the entanglement
measure given by the {\textit{negativity}} \cite{VIDAL,EISERT}.
The negativity, for instance, is determined by the trace norm of
the partially time-reversed state which can be written as
\begin{equation} \label{NEGATIVITY}
 || \vartheta_2\rho ||_1 = \sum_J \frac{\sqrt{2J+1}}{N} \left|
 \sum_K \Theta_{JK} \alpha_K \right|,
\end{equation}
where $|| A ||_1 = {\mathrm{tr}} |A|$ denotes the trace norm of
$A$.

In Refs.~\cite{HORODECKI99,CERF} a necessary separability
criterion, the {\textit{reduction criterion}}, has been introduced
which is based on the positive map defined by $\Lambda
B=I{\mathrm{tr}}B-B$. This criterion is not stronger than the
Peres criterion, but has the important benefit that any state
violating it can be distilled. For $SO(3)$-invariant states the
reduction criterion is equivalent to the inequality based on the
quantum R\'{e}nyi entropy $S_{\infty}$
\cite{HORODECKI96b,HORODECKI99} and to the disorder criterion
\cite{KEMPE}, and takes the form $\frac{1}{N}I\otimes I-\rho\geq
0$. In terms of the parameters $\alpha_J$ this can be expressed
through $\alpha_J\leq \sqrt{2J+1}$. We see explicitly from our
examples that for rotationally invariant states the reduction
criterion is in fact much weaker than the Peres criterion. For
instance, in the case $N=4$ we get from it the conditions
$\alpha_0\leq 1$ and $\alpha_1\leq \sqrt{3}$. The region defined
by these inequalities is much larger than $S_p$ and than the true
set $S_s$ of separable states (see Fig.~\ref{fig3}).

Recently, a necessary criterion for separability has been
developed by Rudolph \cite{RUDOLPH02,RUDOLPH03}, which is known as
{\textit{cross norm}} or {\textit{realignment criterion}}
\cite{CHEN}. This criterion is based on the cross norm of the
states of the tensor product space \cite{RUDOLPH00} and provides
strong conditions for separability. It is generally neither weaker
nor stronger than the PPT criterion. It can detect, however, bound
entanglement. To formulate the cross norm criterion we associate
with any density matrix $\rho = \sum_i C_i \otimes D_i$ a map
$\Phi_{\rho}$ by means of the formula
\begin{equation} \label{DEF-ISO}
 \Phi_{\rho} B = \sum_i C_i {\mathrm{tr}}\{ (\vartheta D_i) B\}.
\end{equation}
For a separable state $\rho$ the corresponding map $\Phi_{\rho}$
is a contraction with respect to the trace norm, i.~e., we have
$|| \Phi_{\rho} ||_1 \leq 1$, which immediately yields a necessary
condition for separability.

The application of the cross norm criterion to rotationally
invariant states leads to an inequality which can again be
expressed entirely in terms of the elements of the matrix
$\Theta$. If the state $\rho$ is given by its parameters
$\alpha_J$ the trace norm of $\Phi_{\rho}$ can be written in a
form analogous to Eq.~(\ref{NEGATIVITY}):
\begin{equation}
 || \Phi_{\rho} ||_1 = \sum_J \frac{\sqrt{2J+1}}{N} \left|
 \sum_K \Theta_{JK} (-1)^K \alpha_K \right| \leq 1.
\end{equation}
This is a general expression for the cross norm criterion of
$SO(3)$-invariant states in any dimension $N$. It allows an
explicit determination of the regions in parameter space
satisfying or violating the criterion. In particular, with the
help of the above formula one immediately evaluates the trace norm
$|| \Phi_{\rho} ||_1$ for the families of the Werner states and of
the isotropic states.

We finally mention that the present results could also find a
number of important applications in the theory of open systems
\cite{TheWork}. The close connection to open system is based on an
isomorphism \cite{JAMIOLKOWSKI} between states $\rho$ on the
tensor product space ${\mathcal{H}}\otimes {\mathcal{H}}$ and
completely positive maps $\Phi$ of operators on ${\mathcal{H}}$.
We define this isomorphism by the relation $\rho=(I\otimes
\Phi)P_0$. Apart from a normalization factor this relation is
equivalent to Eq.~(\ref{DEF-ISO}). It yields a one-to-one
correspondence between the rotationally invariant density matrices
$\rho$ and the completely positive maps $\Phi$ which are
trace-preserving and rotationally invariant. Such maps arise
through the interaction of open systems with isotropic
environments. The isomorphism thus allows one to use the structure
of $S$ in the construction of appropriate representations of
one-parameter families of quantum dynamical maps and to derive the
general form of isotropic non-Markovian quantum processes.

\appendix

\section{Proof of relation (\ref{PJ-QJ})}\label{APP-1}
In the basis of the product states
$|m_1m_2\rangle\equiv|jm_1jm_2\rangle$ the matrix elements of the
operator $\vartheta_2 P_J$ are found to be
\begin{eqnarray} \label{EL-PJ}
 \lefteqn{ \langle m_1m_2|\vartheta_2 P_J|m'_1m'_2\rangle }
 \nonumber \\
 && =(-1)^{2j-m_2-m'_2}\langle m_1,-m'_2|P_J|m'_1,-m_2 \rangle,
\end{eqnarray}
where we have used the definition (\ref{DEF-VARTHETA}) of the
$\vartheta_2$-transformation as well as the matrix elements
(\ref{ELEMENTS-V}) of the unitary matrix $V$ introduced in
Eq.~(\ref{DEF-V}). On the other hand, the definition
(\ref{DEF-TJM}) of the tensor operators and Eq.~(\ref{DERIV-5})
lead to
\begin{eqnarray}
 \langle m|T_{JM}|m'\rangle &=& (-1)^{j-m'}\langle m,-m'|JM\rangle,\\
 \langle m|T^{\dagger}_{JM}|m'\rangle &=& (-1)^{j-m}\langle m',-m|JM\rangle.
\end{eqnarray}
We recall that the matrix elements on the right-hand sides are
vector-coupling coefficients which are taken to be real following
the usual phase conventions. The definitions (\ref{DEF-Q}) and
(\ref{FLIP}) for the operators $Q_J$ and for the flip operator
${\mathbb F}$ yield:
\begin{eqnarray}
 \lefteqn{ \langle m_1m_2|Q_J {\mathbb F}|m'_1m'_2\rangle } \nonumber \\
 && =\sum_{M=-J}^{+J} (-1)^{2j-m_2-m'_2}\langle m_1,-m'_2|JM\rangle
 \langle JM| m'_1,-m_2\rangle \nonumber \\
 && =(-1)^{2j-m_2-m'_2}\langle m_1,-m'_2|P_J|m'_1,-m_2 \rangle.
\end{eqnarray}
Comparing this with (\ref{EL-PJ}) we see that $Q_J {\mathbb F} =
\vartheta_2 P_J$, as claimed.

\section{Proof of inequality (\ref{ALPHA-INEQ})}\label{APP-2}
We take any fixed normalized state $|\varphi^{(2)}\rangle$ and
decompose it with respect to the basis states
$|m\rangle\equiv|jm\rangle$:
\begin{equation} \label{REP-PHI2}
 |\varphi^{(2)}\rangle = c_1\left|+\frac{3}{2}\right\rangle
 +c_2\left|+\frac{1}{2}\right\rangle
 +c_3\left|-\frac{1}{2}\right\rangle
 +c_4\left|-\frac{3}{2}\right\rangle.
\end{equation}
The normalization condition for the amplitudes $c_i$ reads
\begin{equation}
 \sum_{i=1}^4 |c_i|^2 = 1.
\end{equation}
Consider then the operator:
\begin{equation} \label{DEF-A}
 A = \frac{4}{\sqrt{5}}\sum_{M=-2}^{+2}
 T_{2M} |\varphi^{(2)}\rangle\langle\varphi^{(2)}|
 T^{\dagger}_{2M}.
\end{equation}
This operator is obviously Hermitian and positive and we have
$\tilde{\alpha}_2[\varphi^{(1)},\varphi^{(2)}]
=\langle\varphi^{(1)}|A|\varphi^{(1)}\rangle$. It will be
demonstrated below that $|\varphi^{(2)}\rangle$ is an eigenvector
of $A$ corresponding to the eigenvalue $\frac{1}{\sqrt{5}}$:
\begin{equation} \label{EIG-EQ}
 A|\varphi^{(2)}\rangle = \frac{1}{\sqrt{5}}
 |\varphi^{(2)}\rangle.
\end{equation}
This equation implies that $A$ can be written as
\begin{equation}
 A = \tilde{A} + \frac{1}{\sqrt{5}}
 |\varphi^{(2)}\rangle\langle\varphi^{(2)}|,
\end{equation}
where $\tilde{A}$ is again a positive operator. This leads to:
\begin{eqnarray}
 \tilde{\alpha}_2[\varphi^{(1)},\varphi^{(2)}]
 &=& \langle\varphi^{(1)}|\tilde{A}|\varphi^{(1)}\rangle
 + \frac{1}{\sqrt{5}}|\langle\varphi^{(1)}|\varphi^{(2)}\rangle|^2
 \nonumber \\
 &\geq& \frac{1}{\sqrt{5}}\tilde{\alpha}_0[\varphi^{(1)},\varphi^{(2)}],
\end{eqnarray}
which proves the inequality (\ref{ALPHA-INEQ}).

It remains to demonstrate the eigenvector relation (\ref{EIG-EQ}).
To this end, we determine the matrix representation of the
operator $A$ in the basis $|m\rangle$. With the help of the matrix
elements (\ref{DEF-TJM}) of the tensor operators $T_{JM}$ one
finds that $A$ is represented by the matrix
\begin{widetext}
\begin{equation}
 A = \frac{1}{\sqrt{5}} \left[
 \begin{array}{cccc}
 |c_1|^2+2|c_2|^2+2|c_3|^2 & -c_1c_2^*+2c_3c_4^* & -c_1c_3^*-2c_2c_4^* & c_1c_4^* \\
 -c_1^*c_2+2c_3^*c_4 & |c_2|^2+2|c_1|^2+2|c_4|^2 & c_2c_3^* & -c_2c_4^*-2c_1c_3^* \\
 -c_1^*c_3-2c_2^*c_4 & c_2^*c_3 & |c_3|^2+2|c_4|^2+2|c_1|^2 &  -c_3c_4^*+2c_1c_2^* \\
 c_1^*c_4 & -c_2^*c_4-2c_1^*c_3 & -c_3^*c_4+2c_1^*c_2 & |c_4|^2+2|c_3|^2+2|c_2|^2 \\
 \end{array}
 \right].
\end{equation}
\end{widetext}
It is now easy to verify by an explicit calculation that the
vector $c=(c_1,c_2,c_3,c_4)^T$, which represents the state
$|\varphi^{(2)}\rangle$ according to Eq.~(\ref{REP-PHI2}), is an
eigenvector of this matrix corresponding to the eigenvalue
$\frac{1}{\sqrt{5}}$. This concludes the proof.

\end{document}